\documentclass[12pt]{article} 
\usepackage[margin=1in]{geometry} 
\usepackage{amsthm,amssymb, amsfonts, amsmath}
\usepackage{titling, fancyhdr}
\usepackage{enumerate}
\usepackage{xfrac}
\usepackage{units}
\usepackage{wasysym}
\usepackage{titling, framed}
\usepackage{graphicx}
\usepackage{color}
\usepackage{listings}
\usepackage{multirow}
\usepackage{booktabs}
\usepackage{float}
\floatstyle{plaintop}
\restylefloat{table}
\usepackage{cancel}
\usepackage{hyperref}
\usepackage{multicol}
\usepackage{booktabs}
\usepackage{mathtools}
\usepackage[dvipsnames]{xcolor}
\graphicspath{{Figures/}}
\usepackage{amssymb}
\usepackage{amsthm,amssymb, amsfonts, amsmath}
\usepackage{units}
\usepackage{graphicx}
\usepackage{listings}
\usepackage{multirow}
\usepackage{booktabs}
\usepackage{float}
\floatstyle{plaintop}
\restylefloat{table}
\usepackage{booktabs}
\usepackage{enumerate}
\usepackage{caption}
\usepackage{subcaption}
\usepackage{sidecap}
\allowdisplaybreaks

\usepackage{titlesec}
\titleformat*{\section}{\normalsize\bfseries}
\titleformat*{\subsection}{\normalsize\itshape}

\begin{document}
\begin{center}
	\textbf{Measured Nondestructive Assay of $^{237}$Np Using Organic Scintillators and Active Neutron Multiplicity Counting}\linebreak[2]
	\[\]
	Michael Y. Hua$^{1,2}$, Thomas A. Plummer$^{1}$, Jesson D. Hutchinson$^{2}$, George E. McKenzie$^{2}$, Shaun D. Clarke$^{1}$, and Sara A. Pozzi$^{1}$\\
\[\]
	\textit{1. Department of Nuclear Engineering \& Radiological Sciences, University of Michigan, Ann Arbor, MI, USA}\\
	\textit{2. Advanced Nuclear Technology Group, Los Alamos National Laboratory, Los Alamos, NM, USA}
\end{center}

\[\]
\begin{abstract}
	The purpose of nondestructive assay in the context of nuclear safeguards is to precisely verify the declared mass of a sample of nuclear material in a noninhibitive amount of time.  $^{237}$Np is a proliferation concern, and the capacity to efficiently assay samples of it is a missing piece in the verification and safeguards toolbox.  The material is subject to the same safeguards as $^{235}$U, is reportable in gram quantities, and is classified as ``other nuclear material" according to the United States Department of Energy.  Given that 3000 kg of $^{237}$Np is annually produced in the US and the bare sphere critical mass is 40-60 kg, it is desirable to augment the safeguards toolbox with a system capable of distinguishing 10 g of $^{237}$Np in a 20-minute measurement.  One measurement modality is neutron multiplicity counting, which relates the detected multiplicity count rates to the amount of fissionable material.  Prior simulation work shows that an organic scintillator-based multiplicity counter can achieve the design criteria, whereas the flagship $^3$He-based system, the Epithermal Neutron Multiplicity Counter, requires much longer measurement times to achieve the same precision.  In this work, simultaneous measurements of a 6-kg sphere of $^{237}$Np by organic scintillator- and $^3$He-based systems are used to confirm the trends in the simulation study; the organic scintillator-based system achieves 1\% uncertainty in the neutron double multiplicity rate on the order of minutes, while the $^3$He-based system requires days to reach the same precision.  In conclusion, the International Atomic Energy Agency should consider the development and deployment of an organic scintillator-based multiplicity counter. 
\end{abstract}
\pagebreak
\section{Introduction and Motivation}
\indent The purpose of sample assay for nuclear safeguards is to verify operator-declared masses of nuclear material in noninhibitive measurement times~\cite{PANDA,Doyle}.  Nondestructive assay traditionally focuses on special nuclear material; however, $^{237}$Np is also a proliferation concern.  The United States Department of Energy classifies $^{237}$Np as other nuclear material and subjects the isotope to the same safeguards as uranium.  It is desirable to investigate detection systems capable of adequately assaying $^{237}$Np because 3000 kg are annually produced, the capacity to assay the material is a missing piece of the nuclear nonproliferation and safeguards toolbox, and typical characteristics that make an isotope unattractive for use in a nuclear weapon (e.g., heat generation and high spontaneous fission rate) are nearly nonexistent for $^{237}$Np~\cite{mikwa_Np,Goddard2016}.  Previous simulation work compared currently-deployed $^3$He systems to an organic-scintillator-based prototype and concluded that the latter system is capable of assaying $^{237}$Np, whereas the state-of-the-art $^3$He systems were incapable in tenable measurement times~\cite{mikwa_Np}.  The purpose of this work is to confirm the results of the simulation study with measured results.

\section{Measurement Specifications}
\begin{figure}[H]
	\centering
	\includegraphics[width=.5\linewidth]{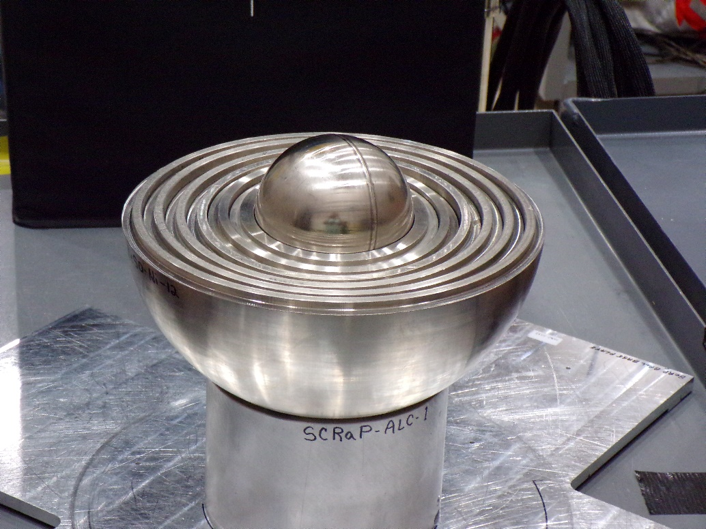}
	\caption{Photograph of the 6-kg $^{237}$Np sphere nested in the bottom half of the nickel hemishells.}
	\label{fig:NpHemi}
\end{figure}

\indent The measurement was performed at the National Criticality Experiments Research Center within the Device Assembly Facility.  The radiation test object (RTO) in this work is a 6-kg sphere of $^{237}$Np, the largest known, single sample of the isotope.  The RTO has impurities that are not uniformly distributed throughout the sample due to mass-separation during the cooling process when the sphere was cast.  The sprue attached to the sphere was removed and chemical analysis was performed, yielding the biased isotopic composition shown in Tab~\ref{tab:isotopics}; note that the weight percentages do not sum to 100\% due to uncertainty, although 100\% is contained within a 95\% confidence interval~\cite{Np_Benchmark}.  The $^{237}$Np sphere is reflected by 7.874 cm of nickel; the nickel comprises several nesting hemishells that assemble into a spherical reflector ~\cite{NeSO1,NeSO2}.  The $^{237}$Np sphere and bottom half of the nickel hemishells are shown in Fig.~\ref{fig:NpHemi}.

\begin{table}[H]
	\centering
	\caption{Isotopic composition of the 6-kg neptunium sphere.}
	\begin{tabular}{lr}
		\toprule
		Isotopes & \multicolumn{1}{l}{Weight Percent} \\
		\midrule
		$^{237}$Np & 98.8000 \\
		$^{233}$U  & 0.0035 \\
		$^{234}$U  & 0.0006 \\
		$^{235}$U  & 0.0276 \\
		$^{236}$U  & 0.0002 \\
		$^{238}$U  & 0.0031 \\
		$^{238}$Pu & 0.0016 \\
		$^{239}$Pu & 0.0314 \\
		$^{240}$Pu & 0.0023 \\
		$^{241}$Pu & 0.0001 \\
		$^{242}$Pu & 0.0003 \\
		$^{241}$Am & 0.0007 \\
		$^{243}$Am & 0.1822 \\
		\bottomrule
		\bottomrule
	\end{tabular}%
	\label{tab:isotopics}%
\end{table}%

\begin{figure}[H]
	\centering
	\begin{subfigure}{.5\textwidth}
		\centering
		\includegraphics[width=.4\linewidth]{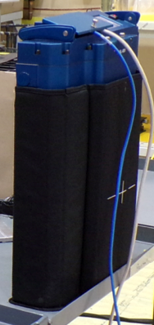}
		\caption{}
		\label{fig:NoMAD}
	\end{subfigure}%
	\begin{subfigure}{.5\textwidth}
		\centering
		\includegraphics[width=.75\linewidth]{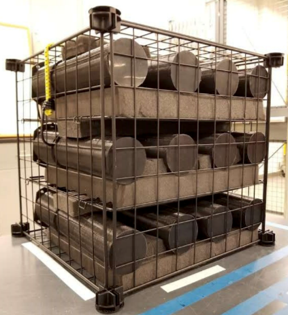}
		\caption{}
		\label{fig:OSCAR}
	\end{subfigure}
	\caption{Photographs of the detection systems; Fig.~\ref{fig:NoMAD} is the Neutron Multiplicity $^3$He Array Detector (NoMAD), whereas Fig.~\ref{fig:OSCAR} is the Organic Scintillator Array (OSCAR) prototype.}
	\label{fig:detectors}
\end{figure}

\begin{figure}[H]
	\centering
	\includegraphics[width=.6\linewidth]{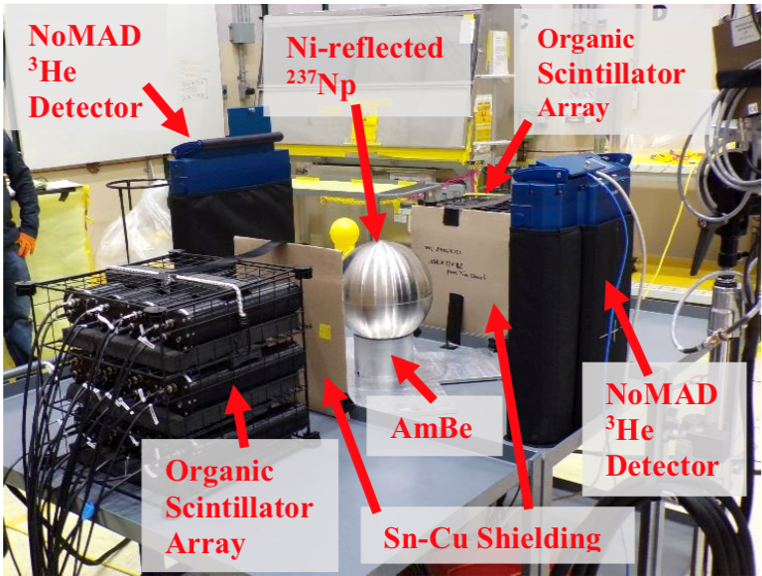}
	\caption{Annotated photo of the measurement setup of the nickel-reflected, 6-kg $^{237}$Np sphere interrogated by AmBe and measured by two NoMAD, $^{3}$He detection systems and two OSCAR, organic-scintillator detection systems.}
	\label{fig:MeasurementSystem}
\end{figure}

Two types of measurement systems are used in this work: the Neutron Multiplicity $^3$He Array Detector (NoMAD) and a prototype of the Organic Scintillator Array (OSCAR) shown in Figs.~\ref{fig:NoMAD} and~\ref{fig:OSCAR}.  The NoMAD detector comprises 15 $^3$He-gas proportional counters embedded in a polyethylene matrix having a minimum clock-tick length of 100 ns, a dead time of $1.5\pm0.3\ \mu$s, and a neutron slowing-down-time of 35-40 $\mu$s (detailed in benchmark-quality in Ref.~\cite{scrap}).  The OSCAR prototype comprises 12, 5.08 cm $\times$ 5.08 cm diameter \textit{trans}-stilbene crystals coupled to photomultiplier tubes~\cite{stilbene,stilbene2,shin_lightoutput}, suspended in powder-coated iron wire meshes and held in place with porous polyurethane foam~\cite{mikwa_RAwOrganics}.  The detectors are powered with high voltage and pulses are digitized with a CAEN v1730 waveform digitizer (16 channels, 500-MHz sampling rate, 14-bit resolution, 2-V dynamic range).  Constant fraction discrimination is used to obtain a time-resolution of 1.34$\pm$0.04 ns, the system has negligible dead time, and each detector is calibrated by adjusting the applied voltage while measuring a $^{137}$Cs source such that 1.6 V-ns pulse integrals correspond to 0.478 MeVee light output.  Two NoMADs and two OSCARs are used in this work; the center-front-faces of each system were 47 cm from the center of the RTO and arranged as shown in Fig.~\ref{fig:MeasurementSystem}.  Note that Fig.~\ref{fig:MeasurementSystem} shows tin-copper graded shielding in front of the OSCARs.  The shielding was used for measurements of plutonium on the same day and were not removed for the measurement of neptunium.  The shields are designed to preferentially shield 60 keV photons from $^{241}$Am and have negligible effect on neutron detection and thus the results of this measurement.

Figure~\ref{fig:MeasurementSystem} indicates that an AmBe source is used to interrogate the $^{237}$Np sphere.  Interrogation is needed because the rate of spontaneous fission in $^{237}$Np is low.  Typically, AmLi sources are used to interrogate $^{235}$U samples; however, neutron from AmLi have a mean energy of 0.59 MeV that is lower than the $^{237}$Np-induced-fission threshold of 0.8 MeV.  Thus AmBe, which has a mean neutron energy of 5.0 MeV and a maximum of approximately 11.0 MeV, is used instead.  The OSCARs measured the RTO assembly for 18 minutes, and the NoMADs measured the RTO assembly for 20 minutes (the last 18 minutes of the NoMAD measurements coincide with the time of OSCAR measurements) ~\cite{PANDA,mikwa_Np,AmLi}.

\section{Data Analysis}
\begin{figure}[H]
	\centering
	\includegraphics[width=.6\linewidth]{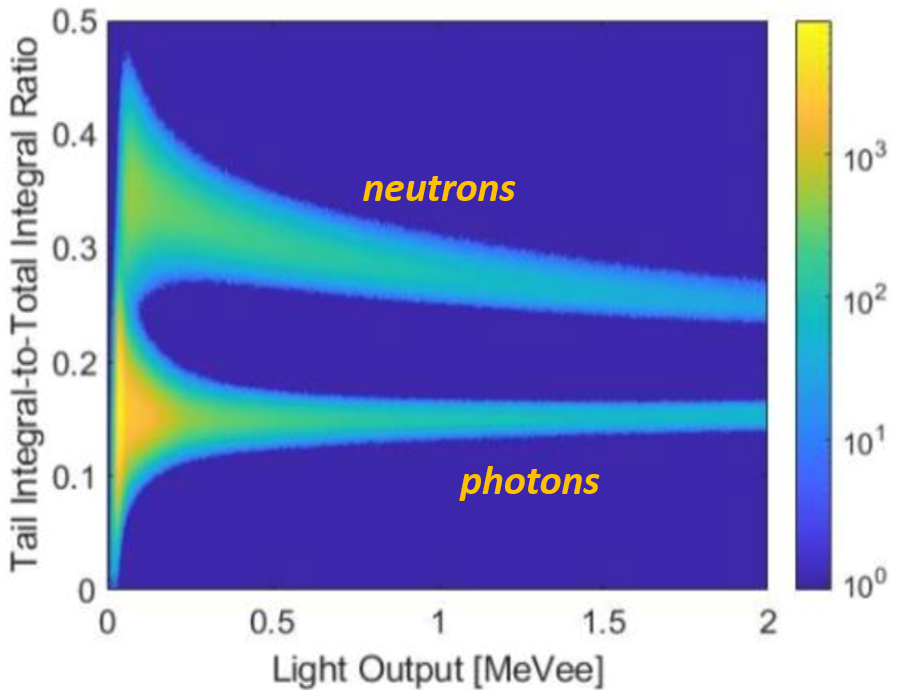}
	\caption{Pulse shape discrimination plot based on a charge integration technique.}
	\label{fig:PSD}
\end{figure}

\indent List-mode data (sorted lists of neutron detection times) are analyzed with factorial moment counting and random trigger intervals; successive intervals of time are inspected for neutron multiplets, which are converted to the neutron double-multiplicity count rate~\cite{hage_cifarelli_factMoment,hage_cifarelli_mult,ensslin_guide}.  The OSCAR uses 100-ns intervals and the NoMAD uses 1-$\mu$s intervals based on Fig. 17 in Ref.~\cite{Jesson_uncertainty}.  In practice, the doubles-multiplicity rate is used with a calibration curve to determine sample mass~\cite{pandendum}.  Statistical uncertainty is propagated analytically~\cite{UQ_jesson}.  One of the NoMAD outputs is the list-mode data.  After initial data pre-processing, the output of the OSCARs is list-mode data including the total pulse integral and the integral of the pulse tail (24 ns after the pulse peak until the end of the pulse).  The integrals are needed to discriminate neutron and photon pulses (since the OSCARs are sensitive to both types of radiation) based on a charge integration technique called pulse shape discrimination (PSD)~\cite{PSD_Slice}.  The PSD plot for this work is shown in Fig.~\ref{fig:PSD}.  The output of the PSD algorithm is list-mode data for neutrons, only.  The list-mode data were combined between measurement systems of the same type.

\section{Results and Discussion}
\begin{figure}[H]
	\centering
	\includegraphics[width=.6\linewidth]{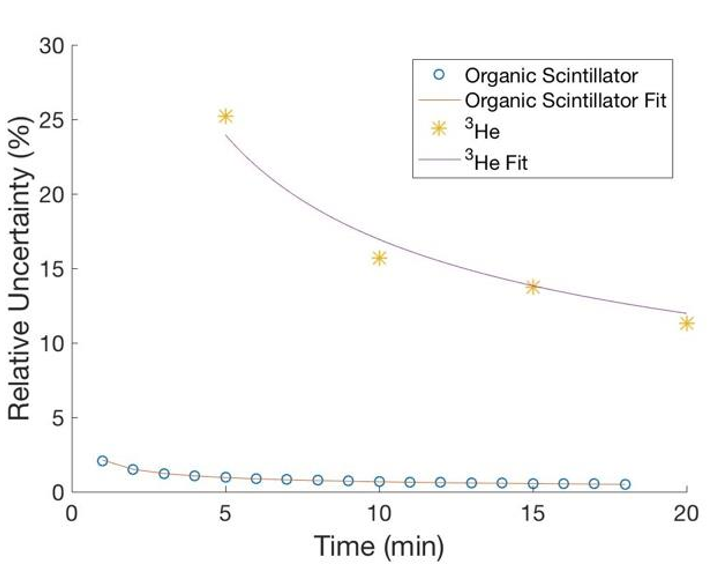}
	\caption{Relative uncertainty as a function of measurement time for the NoMAD, $^3$He system and the OSCAR, organic scintillator system.}
	\label{fig:RelUnc}
\end{figure}

\indent The relative uncertainty was calculated based on the double-multiplicity rate and associated uncertainty for each type of measurement system as a function of measurement time.  The results are shown in Fig.~\ref{fig:RelUnc}.  It is observed that the uncertainty decreases as $(\text{measurement time})^{-1/2}$; the organic scintillator data is fit by
\begin{equation}\label{eq:eq1}
	(\text{Relative Uncertainty})_\text{OSCAR} = 2.131(\text{measurement time})^{-0.5},
\end{equation}
whereas the $^3$He data is fit by
\begin{equation}\label{eq:eq2}
(\text{Relative Uncertainty})_\text{NoMAD} = 53.58(\text{measurement time})^{-0.5}.
\end{equation}
The unit for relative uncertainty is percent and the unit for measurement time is minute in Eqns~\eqref{eq:eq1} and~\eqref{eq:eq2}.  Interpolating for the OSCAR and extrapolating for the NoMAD, the OSCAR requires 4.54 minutes to attain 1\% relative uncertainty while the NoMAD requires approximately 2 days to achieve the same.  Extrapolating for both systems, the OSCAR requires 2.72 seconds to attain 10\% relative uncertainty while the NoMAD requires approximately 28.7 minutes to achieve the same.  Solving Eqns~\eqref{eq:eq1} and~\eqref{eq:eq2} shows that the NoMAD requires a measurement time 632 times longer than the OSCAR.
Note that the uncertainty shown is for statistical uncertainty, only.  Although the NoMAD has greater neutron detection efficiency (13.7 times more based on total neutron counts) than the OSCAR, the total number of inspection intervals/gates is less by four orders of magnitude because the same measurement time is divided by a much larger interval.  The shorter time intervals of the OSCAR are due to the negligible dead time and because time-correlated neutrons are not temporally smeared in moderating material such as the polyethylene in the NoMAD.

A potential source of nonstatistical uncertainty in the OSCAR system is particle misclassification (e.g., classifying a photon as neutron) and neutron crosstalk (a single neutron rendering multiple detections, though this phenomena has been analytically addressed in Ref.~\cite{cross_talk}).  A source of uncertainty for both detection systems is the chemical makeup of the RTO.  Due to the nonuniform distribution of impurities, it is believed that Tab.~\ref{tab:isotopics} is incomprehensive and notably omits curium~\cite{NeSO2}.  Curium has a specific spontaneous fission rate, therein emitting neutron multiplets that are misattributed to the neptunium.  Los Alamos National Laboratory has plans to perform further chemical analysis.

The beryllium in the AmBe interrogation source has a cross section for (n,2n) interactions, which was observed in previous work~\cite{mikwa_Np}.  This correlated signal from AmBe could dominate the desired signal from neptunium from smaller samples and uncertainty in this double rate could define a nonzero, asymptotic uncertainty.  In the former case, higher-order multiplicity rates (such as the triples rate) could be used; however, longer measurement times would be required to attain the same precision as the doubles rate.

\section{Conclusions}
\indent The $^3$He-based NoMAD and organic-scintillator-based OSCAR detection systems, which are similar in form factor, are compared in their capacity to assay a 6-kg sphere of $^{237}$Np by way of multiplicity counting.  The systems are principally compared on precision and the time it takes to achieve the same relative uncertainty in the double-multiplicity count rate; the NoMAD is 632 times slower than the OSCAR.  Besides relative comparisons, the OSCAR can achieve excellent precision in under five minutes and moderate precision in under three seconds, making the prototype highly reasonable for field deployment.  The reduced measurement times will, for example, enable inspectors to inspect more samples in lieu of randomly selecting a hopefully representative subset.  Reduced measurement times will also reduce procedural and operational costs.  Thus, it is recommended that organic-scintillator-based systems be considered as upgrades to currently deployed $^3$He systems.  

The OSCAR prototype has not been optimized for efficiency, yet the rapid-assay capability lends the system to applications beyond verification of operator-declared masses.  For example, the OSCAR could be reconfigured to affix to a pipe and assay moving material as it passes, depending on the mass flow rate.  Future work includes simulated studies of pipe-monitoring applications, efficiency optimizations, and testing in field-like conditions.

\section*{Acknowledgments}
This work was partially supported by the  National Science Foundation Graduate Research Fellowship under Grant No. DGE-1256260, the Consortium for Verification Technology under Department of Energy National Nuclear Security Administration award number DE-NA0002534, the Consortium for Monitoring, Technology, and Verification under Department of Energy National Nuclear Security Administration award number DE-NA0003920, and the DOE Nuclear Criticality Safety Program, funded and managed by the National Nuclear Security Administration for the Department of Energy.  Any opinion, findings, and conclusion or recommendations expressed in this material are those of the authors and do not necessarily reflect the views of any funding organization. 

\bibliographystyle{ans_js}
\bibliography{bibliography}
\end{document}